\documentclass[preprint, showpacs,showkeys,preprintnumbers,amsmath,amssymb]{revtex4}
\usepackage{graphicx,amscd,amsmath,amssymb,float}
\usepackage{dcolumn}
\usepackage{tabularx}
\usepackage{bm}

\usepackage{color}

\begin{document}
\title{Adsorption of Sugars on Al- and Ga-doped Boron Nitride Surfaces: A Computational Study} 
\author{Ahmed A.\ Darwish $^{1,3}$}\
\author{Mohamed M.\ Fadlallah$^{2,4}$}\
\author{Ashraf Badawi$^{1}$}\
\author{Ahmed A.\ Maarouf $^{2,5,6}$}\
\affiliation{
$^1$ Center for Nanotechnology, Zewail City of Science and Technology, Giza 12588, Egypt\\
$^2$ Center for Fundamental Physics, Zewail City of Science and Technology, Giza 12588, Egypt\\
$^3$ Department of Nuclear \& Radiation Engineering, Faculty of Engineering, Alexandria University, Alexandria, Egypt\\
$^4$ Department of Physics, Faculty of Science, Benha University, Benha, Egypt\\
$^5$ Egypt Nanotechnology Center, Cairo University, Giza 12613, Egypt\\
$^6$ Department of Physics, Faculty of Science, Cairo University, Giza 12613, Egypt}

\date{\today}
\begin{abstract}
Molecular adsorption on surfaces is a key element for many applications, including sensing and catalysis. Non-invasive sugar sensing 
has been an active area of research due to its importance to diabetes care. The adsorption of sugars on a template surface study is 
at the heart of matter. Here, we study doped hexagonal boron nitride sheets ($h$-BNNs) as adsorbing and sensing template for glucose and glucosamine. Using first principles calculations, we find that the adsorption of glucose and glucosamine on $h$-BNNs is significantly enhanced by the substitutional doping of the sheet with Al and Ga. Including long range van der Waals corrections gives adsorption energies of about 2 eV. In addition to the charge transfer occurring between glucose and the Al/Ga-doped BN sheets, the adsorption alters the size of the band gap, allowing for optical detection of adsorption. We also find that Al-doped boron nitride sheet is better than Ga-nitride sheet to enhance the adsorption energy of glucose and glucosamine. The results of our work can be potentially utilized when designing support templates for glucose and glucosamine.
\end{abstract}


\maketitle
 
\section{Introduction}

Molecular adsorption plays an important role in many applications, such as catalysis \cite{ptongraph1, cat2, cat3}, molecular sensing \cite{Schedin2007, Yue2013}, photovoltaics \cite{pv1, pv2}, and chemical separation \cite{chemsep1, chemsep2}. The adsorption process typically involves changes in the physical or chemical properties of the adsorbed molecule and the adsorbing surface. For example, the change in the electronic properties of some graphene structures has been used for molecular sensing \cite{graphsense1, graphsense2,maaroufpatent}. Modelling these systems is important for a complete understanding and prediction of their properties so that sensor paradigms can be designed.

Hexagonal boron nitride structures have been considered for many applications, including nanoelectronics  \cite{kdh,khl}. In the area nanobiology, the substance's biocompatiblity has made it useful in drug delivery and biosensing \cite{msa,Weng,jwl}. For example glucose biosensors using boron nitride nanotubes (BNNTs) have shown high sensitivity and stability, good reproducibility, anti-interference ability, especially excellent acid stability and heat resistance \cite{jwl}. Theoretical studies have also shown that BNNTs can be used for sensing molecules such as methane, paracetamol, nitophenol, and chitosan \cite{Chowdhury,Soltani,Anurag,Khaled,Darvish2010,Chigo,Ahmadi,Rodriguez}. 

Fabricated boron nitride nanostructures typically have adatoms \cite{ov,jh}, vacancies \cite{ol}, and Stone Wales defects \cite{wch,ykch}. Substitutional doping of BN structures with aluminum atoms has been shown to enhance the adsorption of polar molecules \cite{sm,sn}. This is mainly due to the difference of electronegativity between the boron and aluminum, which creates an electrostatically distinct location. Therefore, an atom scale investigation into the principles underlying the binding to BN structure and observable changes to their electronic properties is of interest.

In this work, we study glucose and glucosamine adsorption onto pristine and doped hexagonal boron nitride nanosheets ($h$-BNNs) using first principles calculations. We substitutionally dope the $h$-BNNs with Al and Ga at the B site. The effect of the glucose (G) and glucosamine (Gl) orientation on their adsorption onto the doped $h$-BNNs, as well as their effect on the electronic properties of the Al- and Ga-doped $h$-BNNs are investigated. We also study the influence of the dopant concentration on the adsorption properties of G and Gl, as well as on the electronic properties of the doped sheets. We show that the proposed doped $h$-BNNs systems demonstrate reasonable differentiability between G and Gl, and can thus act as effective sensors.

\section{Computational Method}
Density-functional theory calculations were carried out by using the Quantum Espresso (QE) package \cite{esp}. Ultrasoft pseudopotentials, 
the Perdew-Burke-Ernzerhof (PBE) generalized gradient approximation (GGA) \cite {gga1,gga2} have been employed and van der Waals interactions 
were considered within the semi-empirical DFT-D2 \cite{Dion} and the first-principles vdW-DF methods \cite{grimme}. An energy cutoff of 45 Ry, and a $9 \times 9 \times 1$ Monkhorst-Pack $k$-point grid were used \cite {Mon}. The supercell size was such that the vacuum separation in the z-direction was 12 $\mathrm{\AA}$. The charge transfer was determined using the L{\"o}wdin projection of charges. All systems were structural relaxation till the forces were less than 0.002 Ry/Bohr.

\section{Results and Discussion}

We study the adsorption of G and Gl onto pristine and doped $h$-BNNs. Since there is more than one possible adsorption site in each molecule, we first determine the orientation of the target molecule with respect to the $h$-BNNs that has the lowest energy, then determine the properties of the system in the preferred orientation. We find that the preferred orientation is typically lower in energy than other orientations by about 10 $k_B T$. In order to address the effect of the dopant concentration on the electronic properties of the system, three supercell sizes with one dopant atom per supercell have been considered: $4 \times 4$, $5 \times 5$, and $6 \times 6$  $h$-BNNs unit cells, which correspond to a doping capacity of about $3$\%, $2$\%, and $1.4$\%, respectively. For each of these supercells, doping with Al and Ga is considered. 

The adsorption energy of the adsorbed molecule ($E_{d}$) onto the $h$-BNNs has been calculated using the following formula:
\begin{equation}
E_{\mathrm{d}} = E_{\mathrm{molecule+x\mbox{-}BN}} - E_{\mathrm{x\mbox{-}BN}} - E_{\mathrm{molecule}} ,
\end{equation}
where, $E_{\mathrm{molecule+x\mbox{-}BN}}$ is the total energy of the target-$h$-BNNs system, $E_{\mathrm{x\mbox{-}BN}}$ is the total energy of the doped sheet, and $E_{\mathrm{molecule}}$ is the total energy of the isolated molecule.

\subsection{Adsorption of glucose and glucosamine onto pristine $h$-BNNs }

\begin{figure}[H]
\begin{center}
\includegraphics[width=0.4\textwidth]{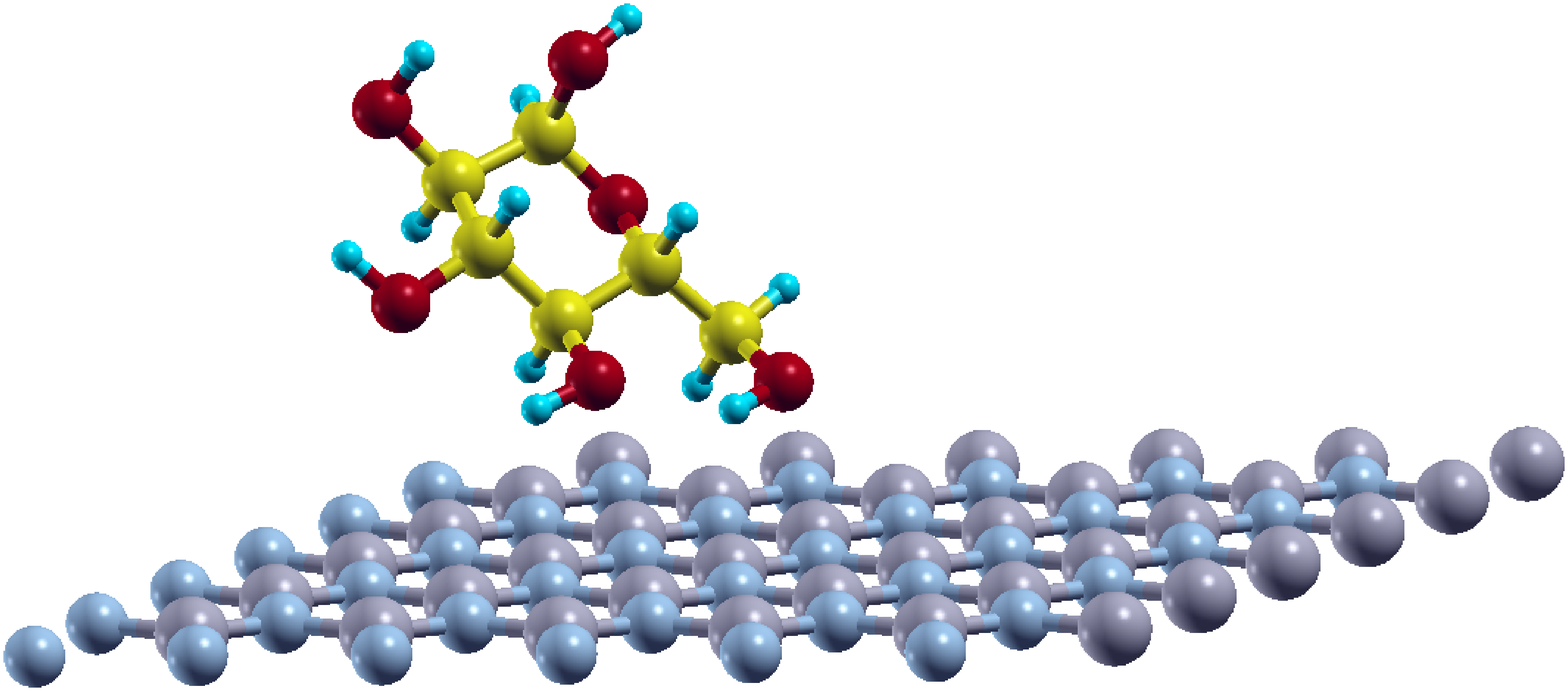}~\hspace{1mm}
\includegraphics[width=0.4\textwidth]{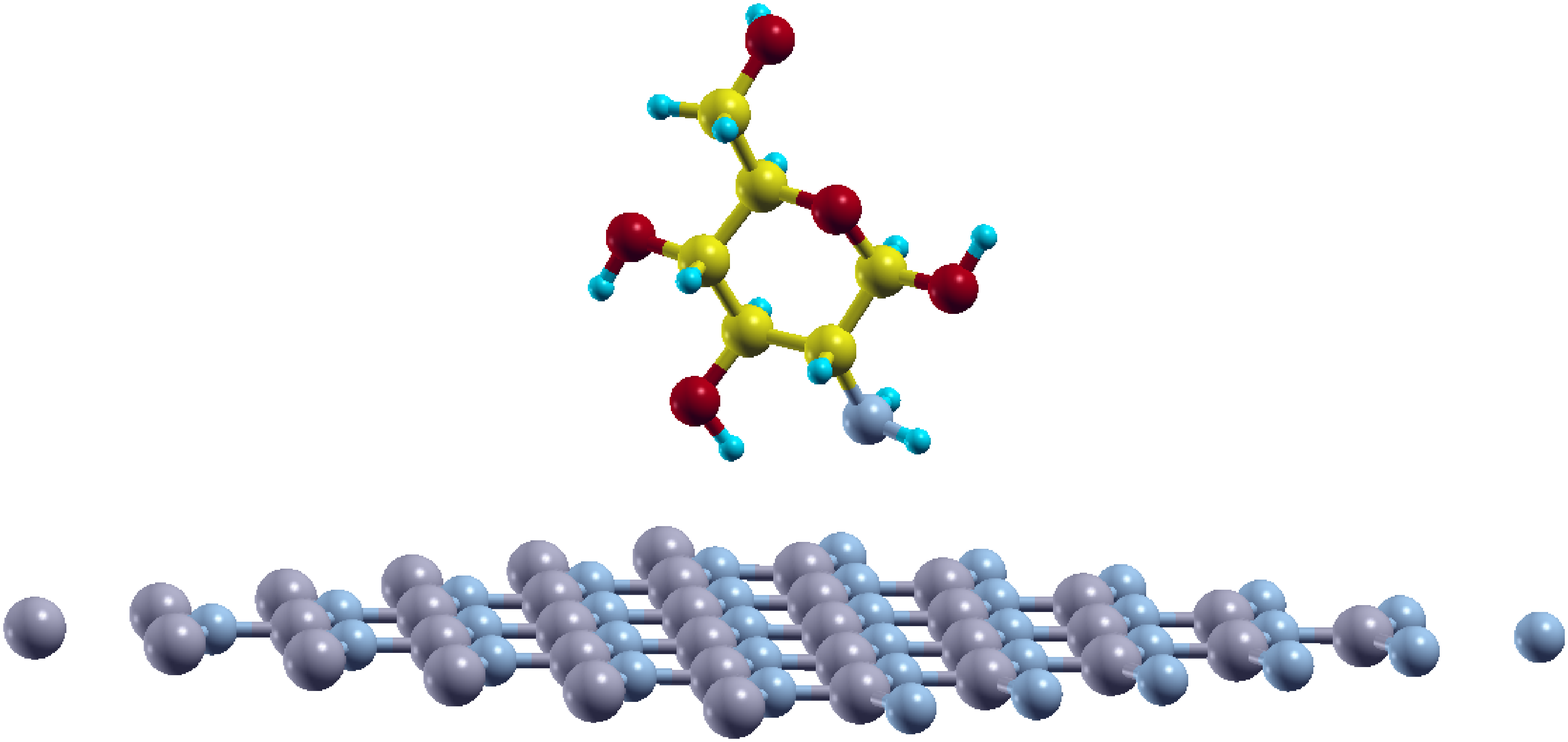}
\caption{(Color online) Adsorption of glucose (left) and glucosamine (right) molecules on top of pristine BNNs (left). The color scheme is as follows: Carbon is in yellow, oxygen is in red, nitrogen is in pale blue, boron is in gray, and hydrogen in light blue.}
\label{gl-pr}
\end{center}
\end{figure}

Structural relaxation of the pristine $h$-BNNs gives us a B-N bond length of $1.45$ $\mathrm{\AA}$, and an electronic band gap of $4.2$ eV, which fall within theoretical and experimental ranges reported in the literature \cite{sn,jhw,na,ss,hs,xb,ll,mt,ana}. Various configurations of G onto $h$-BNNs (G@$h$-BNNs) and Gl onto $h$-BNNs (Gl@$h$-BNNs) are structurally optimized, and their electronic properties are determined.  Figure \ref{gl-pr}(left) shows the lowest energy configuration of G on a $6 \times 6$ supercell of $h$-BNNs supercell. The O-B distance is found to be $3.6$ $\mathrm{\AA}$. The adsorption energy is found to be very weak, $~0.05$ eV. Using a L{\"o}wdin charge analysis, no significant charge transfer occurs between the G molecule and the $h$-BNN. Similar results are obtained for the adsorption of Gl onto $h$-BNNs (Fig. \ref{gl-pr}(right)).

\subsection{Al- and Ga-doped $h$-BNNs }

\begin{figure}[H]
\begin{center} 
\includegraphics[width=0.8\textwidth]{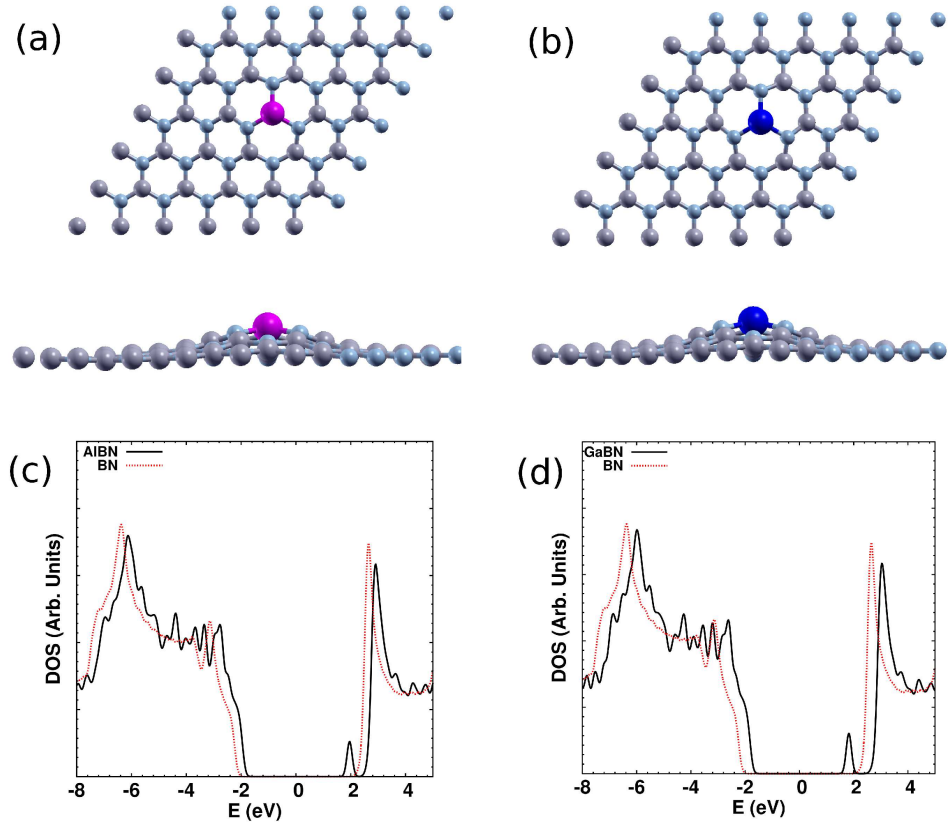}
\caption{(Color online) (a) Al-doped $6 \times 6$ BN system ($Al$-BNNs). The Al atom is in purple. (b) Projected density of states of the $Al$-BNNs system. (c) Ga-doped $6 \times 6$ BN system ($Ga$-BNNs). The Ga atom is in dark blue. (d) Projected density of states of the $Ga$-BNNs system.}
\label{AlGaBN}
\end{center}
\end{figure}

To increase the activity of the $h$-BNNs, one can substitutionally dope its lattice with various elements. To maintain the same coordination, trivalent elements, such as Al and Ga, are preferred. Figure \ref{AlGaBN}a shows the structurally relaxed Al-doped $h$-BNNs (Al-$h$-BNNs), where a B atom is replaced by an Al atom, corresponding to a doping density of $1.4$\%, whereas Fig. \ref{AlGaBN}b shows the corresponding Ga-$h$-BNNs system. Since Al and Ga have atomic radii larger than B ($R_{Al}=1.18 \mathrm{\AA}$, $R_{Ga}=1.26 \mathrm{\AA}$, and $R_{B}=0.87 \mathrm{\AA}$), they protrude out of sheet, with a Al-N bond length of $1.735 \mathrm{\AA}$ and a N-Al-N angle of $117.2$\r{}, and a Ga-N bond length of $1.78 \mathrm{\AA}$ and a N-Ga-N angle of $116.22$\r{}. The Al and Ga bonds distorted the hexagonal structure of $h$-BNNs, which is in agreement of what has been reported in the case of Al/Ga-doped boron nitride, graphene, and carbon nanotubes systems \cite{sn,rw,zm}. 

In Fig. \ref{AlGaBN}c, we show the projected density of states (PDOS) of the Al-$h$-BNNs system and compare it to that of the pristine $h$-BNNs. The band gap of the Al-$h$-BNNs system is decreased to $3.7$ eV due to the Al state that now exists below the conduction band edge of $h$-BNNs. L{\"o}wdin charge analysis shows that the Al atom has a charge of $+1.4e$ and the three neighboring nitrogen atoms have a charge of $-0.7e$. In the pristine $h$-BNNs, a boron atom has a charge of $+0.5 e$, while a nitrogen atom has a charge of $-0.5e$. Fig. \ref{AlGaBN}d shows that the Ga-$h$-BNNs system has a qualitatively similar structure to the Al-$h$-BNNs one, but with a gap of $3.64$ eV. The charge on the Ga atom is found to be $+1e$, with a charge of $-0.6e$ on its three nitrogen nearest neighbors. The higher charging of Aluminum can be attributed to its lower electronegative, as compared to Ga. Therefore, the presence of the Al and Ga dopants creates a local distortion of the bipolar charge anatomy of the BN lattice.

\subsection{ Adsorption of glucose onto Al- and Ga-doped $h$-BNNs } 
 
\begin{figure}[H]
\begin{center}
\includegraphics[width=0.8\textwidth]{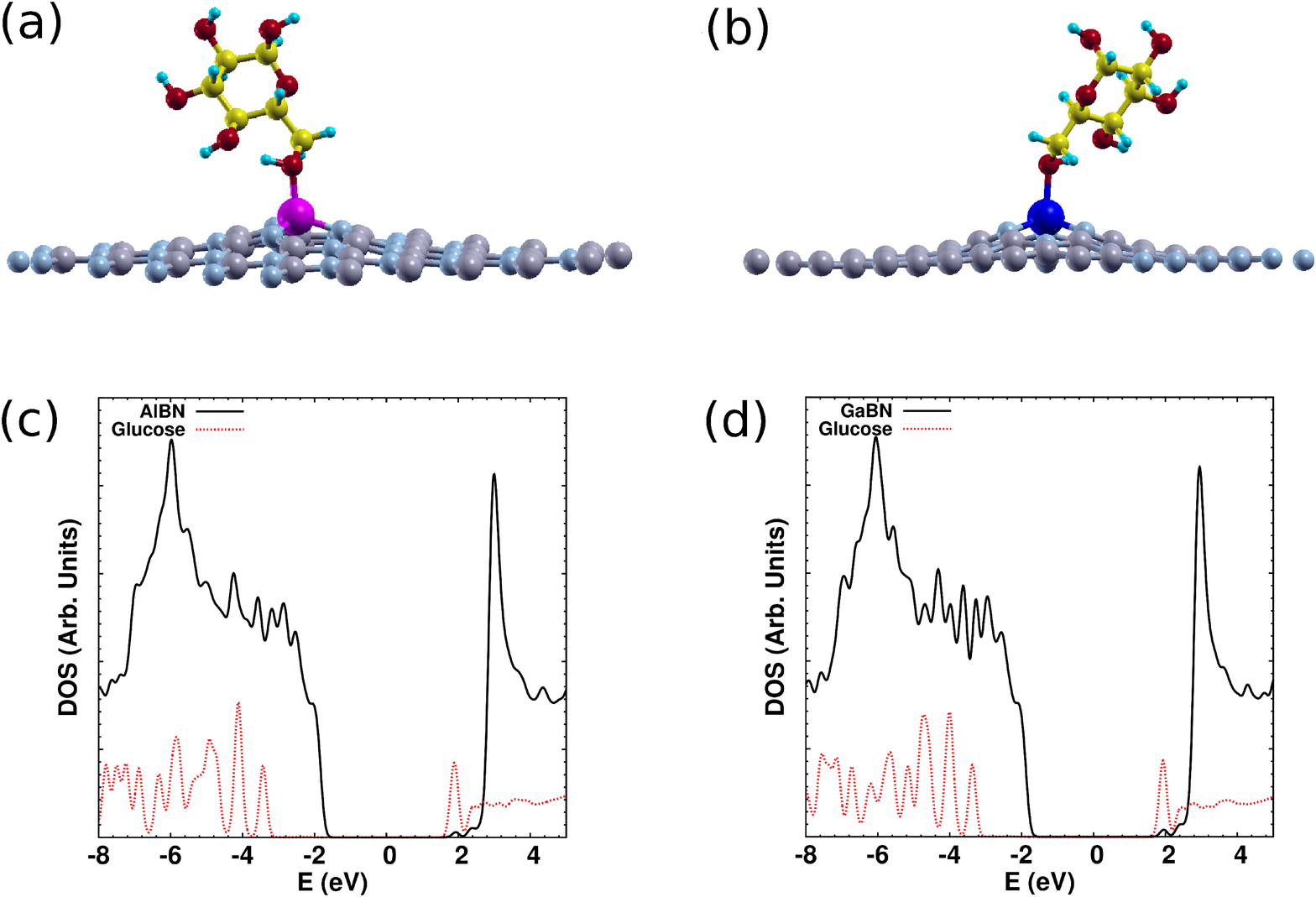}
\caption{(Color online) (a) Glucose on top of Al-doped $6 \times 6$ BN system (G@Al-$h$-BNNs). (b) Glucose on top of Ga-doped $6 \times 6$ BN system (G@Ga-$h$-BNNs). (c) Projected density of states of the G@Al-$h$-BNNs system. (d) Projected density of states of the G@Ga-$h$-BNNs system.}
\label{G_AlBN}
\end{center}
\end{figure}

Next we address the results for the adsorption of the G and the Gl molecules onto the Al- and Ga-doped $6 \times 6$ supercell of $h$-BNNs, at the Al and Ga sites (Fig. \ref{G_AlBN}). We first consider the G-molecule. We find that the orientation with the lowest energy is with the oxygen atom of the hydroxyl group adsorbed onto the Al site, with an O-Al bond length of $1.90 \mathrm{\AA}$, as we show in Fig. \ref{G_AlBN}a. The G molecule pulls the Al atom up, extending the Al-N bond from $1.74 \mathrm{\AA}$ to $1.82 \mathrm{\AA}$, and decreasing the N-Al-N angle from 117\r{} to 110\r{}. These changes bring the Al orbitals closer to an {\it sp}$^{3}$ configuration \cite{Ahmadi}. We find the adsorption energy of G onto Al-$h$-BNNs to be $1.28$ eV, which is significantly higher than onto the pristine $h$-BNNs.  L{\"o}wdin charge analysis shows that about $0.07 e$ is transferred from the oxygen atom to the Al atom. The overall charge transferred between the G molecule and the Al-$h$-BNNs is $0.22 e$. The adsorption process results in a slight elongation of the bond C-O bond of glucose closest to the Al atom from $1.42 \mathrm{\AA}$ to $1.46 \mathrm{\AA}$.

In Fig. \ref{G_AlBN}b, we show the G@Ga-$h$-BNNs system. The structural details of the adsorption are qualitatively similar to those of Al, with a O-Ga bond length of $2.1 \mathrm{\AA}$. The adsorption increases the Ga-N distance from $1.78$ to $1.83 \mathrm{\AA}$. The adsorption energy of G onto Ga-$h$-BNNs is $0.79$ eV. The local charge transferred from the oxygen atom to the Ga atom is $0.05e$, while $0.21e$ is transferred from the G-molecule and the Ga-$h$-BNNs.

In Fig. \ref{G_AlBN}c, we show the PDOS of the G@Al-$h$-BNNs system. The HOMO of the glucose falls below the valence band of the Al-$h$-BNNs, whereas its LUMO falls below the bottom of the conduction band, causing a decrease of the band gap from $3.9$ to $3.6$ eV. This reduction in the band gap can be utilized for sensing glucose. Analysis of the states in the vicinity of the gap shows that the Al state that was previously present in the gap has now shifted to other parts of the energy spectrum due to the coupling with the oxygen states. 

The PDOS of the G@Ga-$h$-BNNs system is shown in Fig. \ref{G_AlBN}d . As in the Al-$h$-BNNs case, the HOMO of the G falls below the valence band of the Ga-$h$-BNNs. The G LUMO falls closer to the bottom of the conduction band. Coupling between Ga and the oxygen states moves the Ga 3$s$ state to other parts of the energy spectrum. 

\subsection{ Adsorption of glucosamine onto Al- and Ga-doped $h$-BNNs } 

\begin{figure}[H] 
\begin{center}
\includegraphics[width=0.8\textwidth]{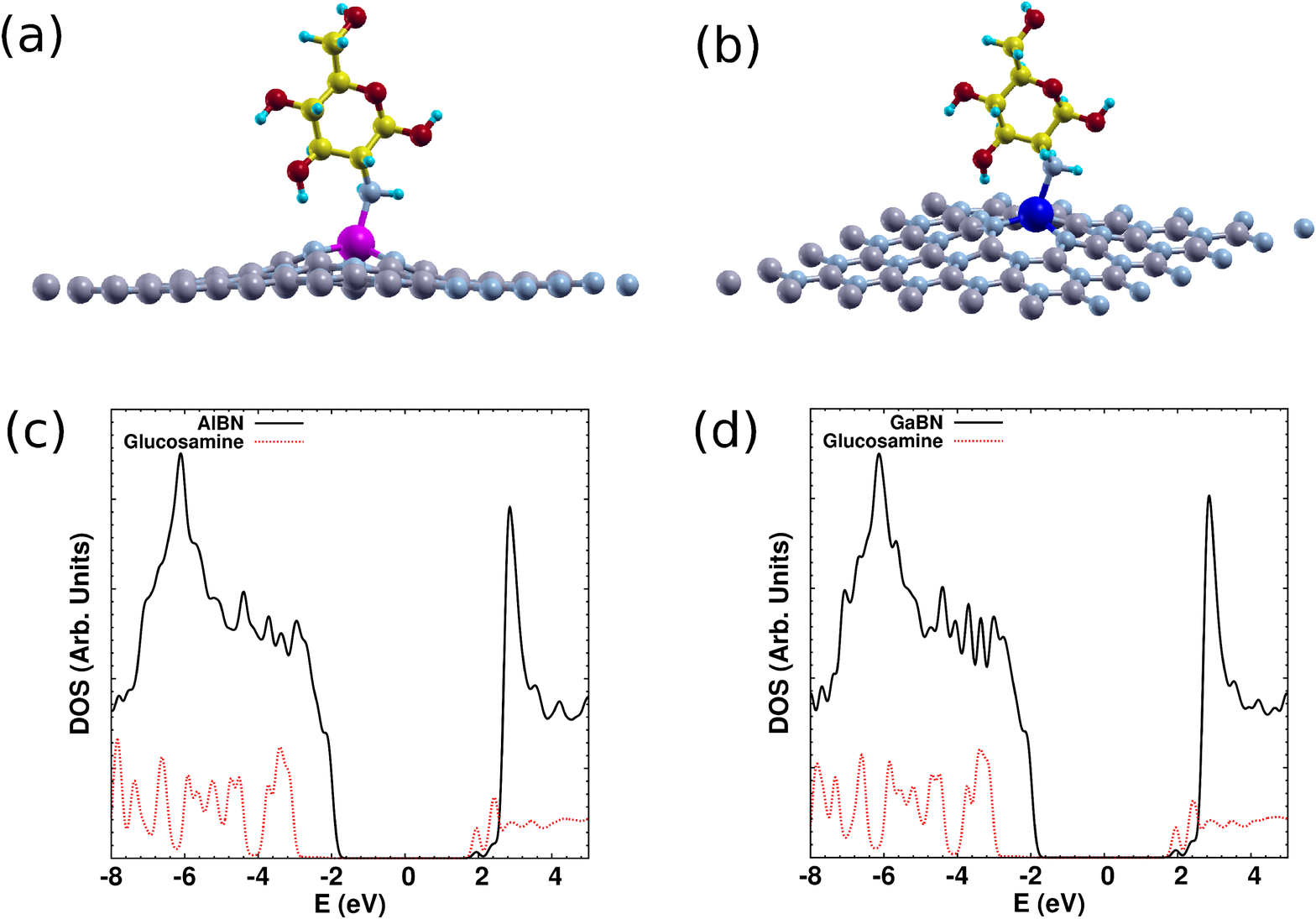}
\caption{(Color online) (a) Glucosamine on top of Al-doped $6 \times 6$ BN system (Gl@Al-$h$-BNNs). (b) Glucosamine on top of Ga-doped $6 \times 6$ BN system (Gl@Ga-$h$-BNNs). (c) Projected density of states of the Gl@Al-$h$-BNNs system. (d) Projected density of states of the Gl@Ga-$h$-BNNs system.}
\label{Gl_AlBN}
\end{center}
\end{figure}

We now discuss our results for the adsorption of glucosamine on a $6\times 6$ supercell of Al- and Ga-doped $h$-BNNs systems. By studying various orientations of the Gl molecule on the sheet, we find that the lowest energy configuration is that with the glucosamine amine group attached to the Al or Ga atom (Fig. \ref{Gl_AlBN}a,b). The distance between the amine group nitrogen and the Al atom is $2.0 \mathrm{\AA}$, and slightly larger between the amine group nitrogen and the Ga atom.  The $sp^3$ bonding nature of the Al and Ga atoms is increased due to the adsorption of glucosamine, and the subsequent elongation of the $h$-BNNs Al-N and the Ga-N bonds to $1.77 \mathrm{\AA}$ and $1.83 \mathrm{\AA}$, respectively. The L{\"o}wdin charge analysis indicates that the nitrogen of the amine group loses charge locally at the adsorption site; with $0.08e$ transferred to the Al atom in the case of Gl@Al-$h$-BNNs, and $0.1e$ transferred to the Ga atom in the case of Gl@Ga-$h$-BNNs. The adsorption energy of the Gl molecule onto Al-$h$-BNNs is $1.29$ eV, and is $0.94$ eV onto Ga-$h$-BNNs.

In Fig. \ref{Gl_AlBN}c and \ref{Gl_AlBN}d, we show the PDOS of the Gl@Al-$h$-BNNs  and Gl@Ga-$h$-BNNs $6 \times 6$ systems. The HOMO of the Gl-molecule falls below the valence band edge of the Al and G doped $h$-BNNs, while its LUMO falls below the conduction band edge, causing a change in the gap. The adsorption leads to the coupling between the amine group nitrogen and the Al and Ga atoms, sending Al 3$s$ state and Ga $3s$ state away from the gap to other locations in the energy spectrum. In the two doped systems, glucosamine adsorption slightly increases the gap to $3.85$ eV.
 
\subsection{ Effect of doping concentration } 

In the systems discussed so far, we had one dopant atom per $6 \times 6$ supecell, which amounts to a doping percentage of $1.4$\%. To investigate the effect of doping concentration on the adsorption of glucose on Al- and Ga-doped $h$-BNNs, we repeated the aforementioned set of calculations on $5 \times 5$ and $4 \times 4$ $h$-BNNs systems, which correspond to a doping percentage of about $2$\% and $3$\%, respectively. 

\begin{figure}[H]
\begin{center}
\includegraphics[width=0.4\textwidth]{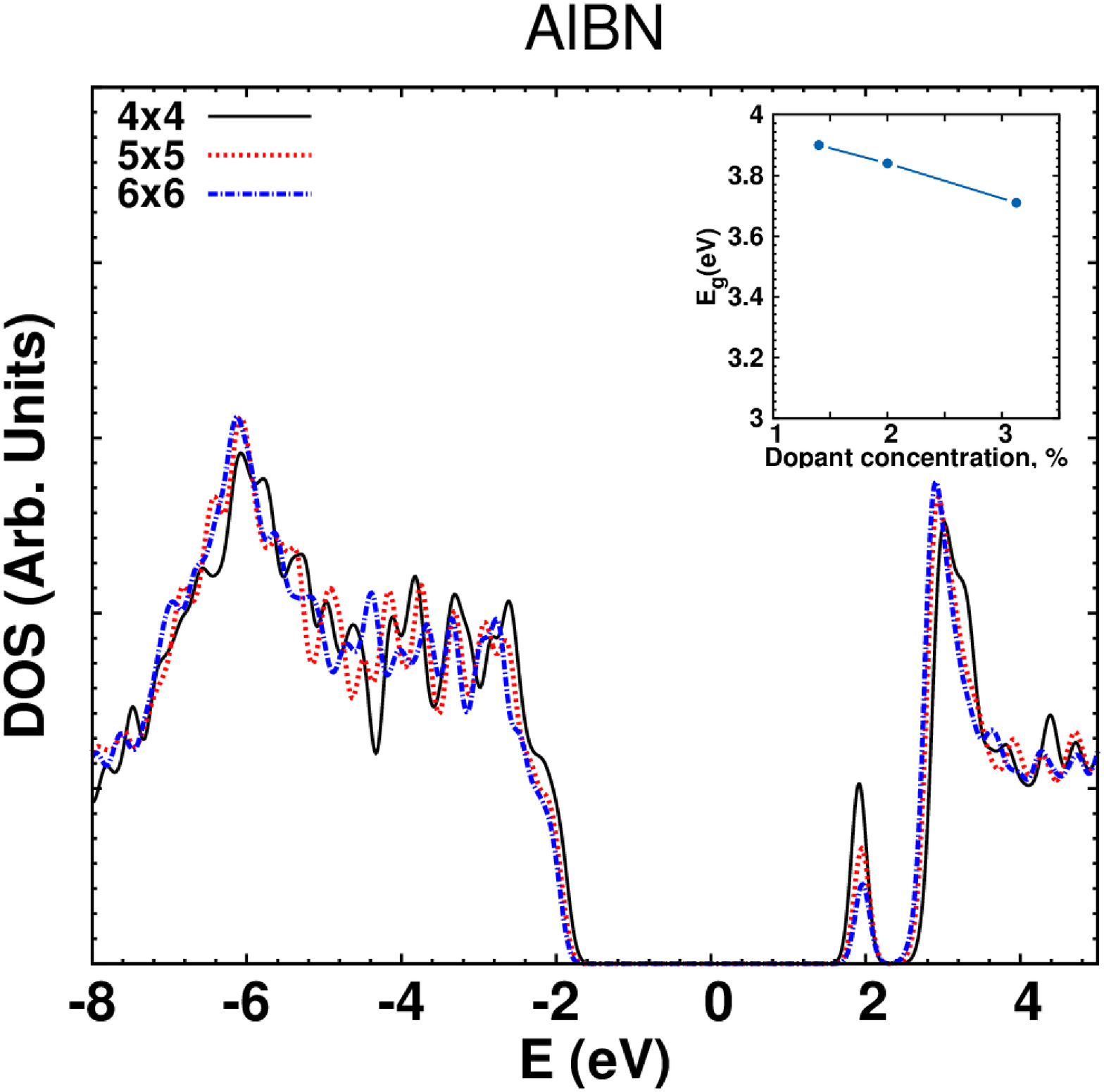}~\hspace{4mm}
\includegraphics[width=0.4\textwidth]{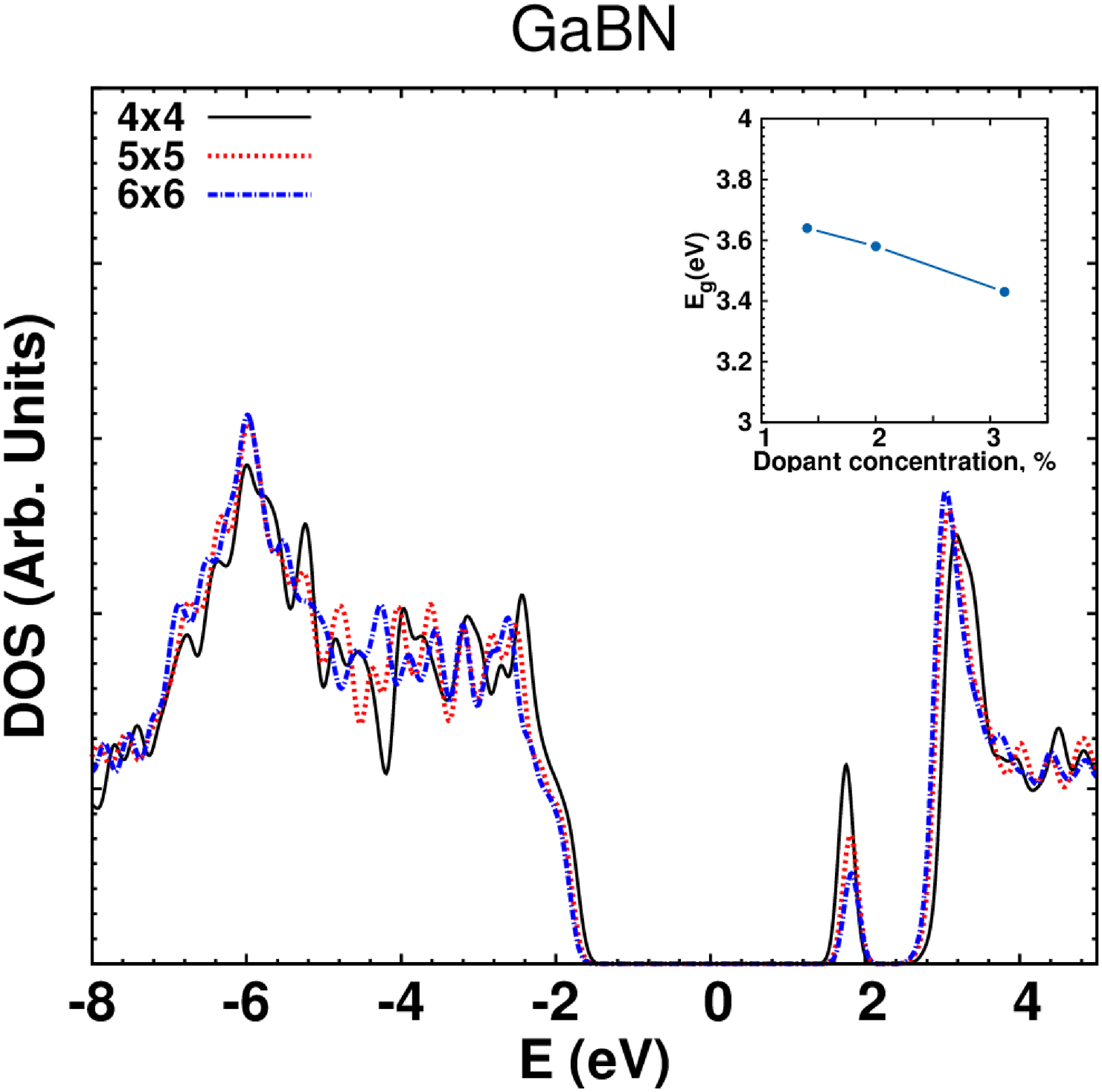}
\caption{(Color online) (Left) PDOS of the Al-$h$-BNNs at various doping concentrations. (Right) PDOS of the Ga-$h$-BNNs at various doping concentrations (right). The insets show the dependence of the energy gaps of the Al ad Ga doped $h$-BNNs structures on the doping concentration.}
\label{dopingconcAlGa}
\end{center}
\end{figure}

We first consider the Al- and Ga-doped $h$-BNNs systems. Figure \ref{dopingconcAlGa} (left) shows the PDOS of Al-$h$-BNNs system for the three considered doping percentages, indicating its minor effect on the electronic properties. The inset shows the band gap slightly decreases with dopant concentration. Figure \ref{dopingconcAlGa} (right) and its inset lead to a similar conclusion for the Ga-$h$-BNNs system.

\begin{figure}[H]
\begin{center}
\includegraphics[width=0.4\textwidth]{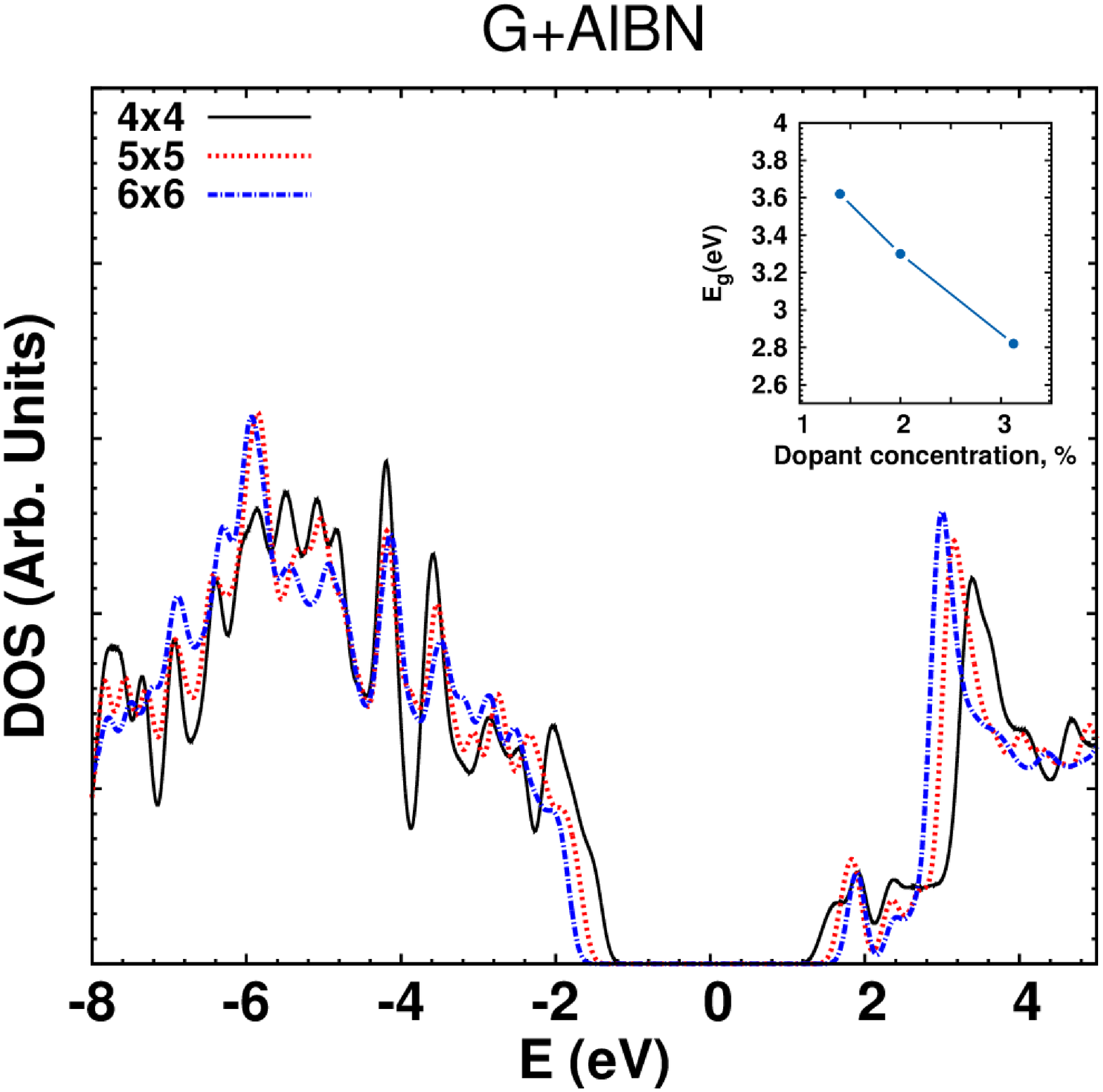}~\hspace{4mm}
\includegraphics[width=0.4\textwidth]{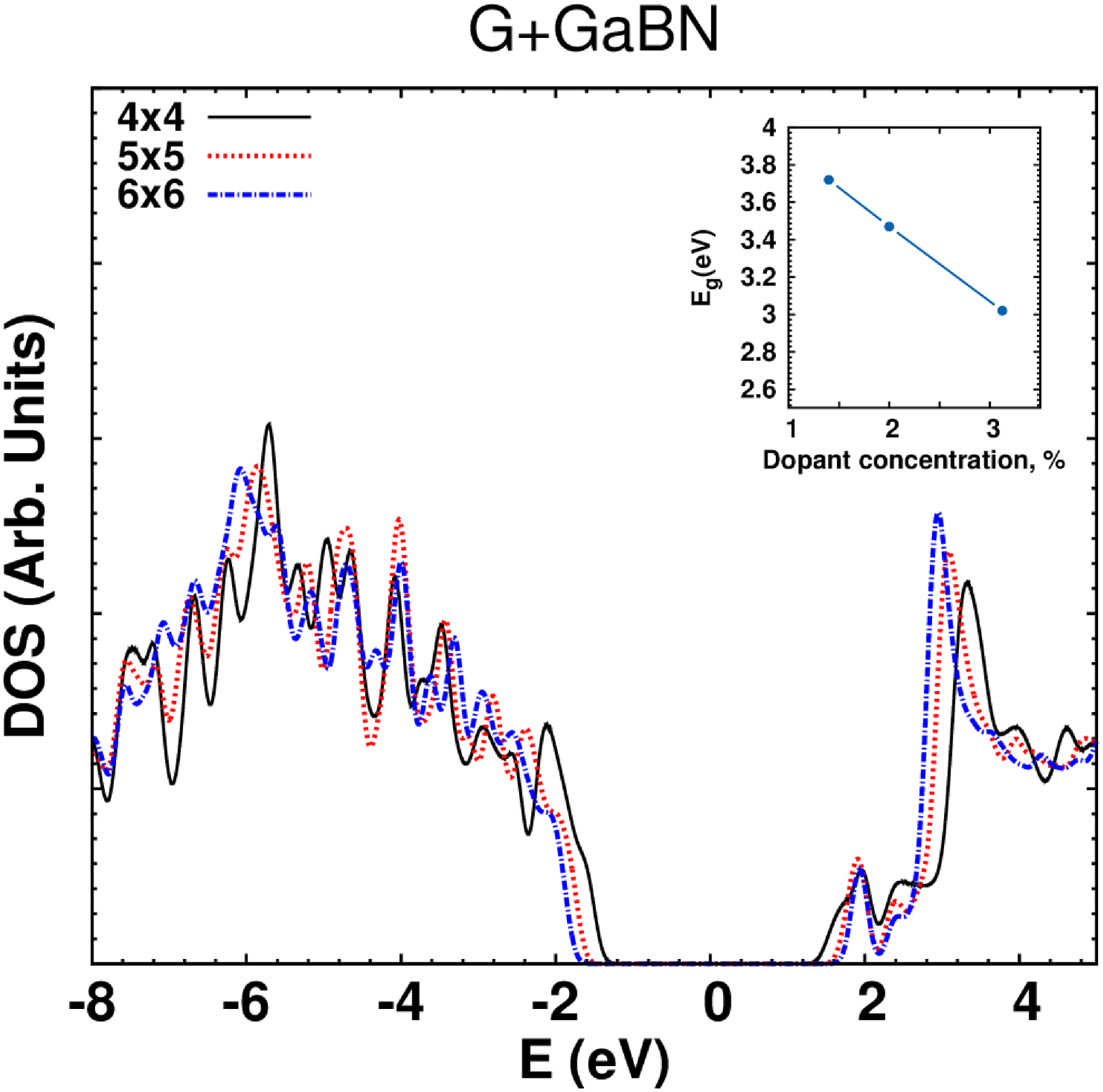}
\caption{(Color online) DOS of glucose adsorbed onto Al-doped $h$-BNNs, G@Al-$h$-BNNs at various doping concentrations (left). DOS of glucose adsorbed onto Ga-doped $h$-BNNs, G@Ga-$h$-BNNs at various doping concentrations. Insets show the dependence of the energy gaps of both systems on the doping concentration (right).}
\label{dopingconcGlucose_AlGa}
\end{center}
\end{figure}

We now turn to the effect of doping concentration on the adsorption of glucose onto Al-$h$-BNNs and Ga-$h$-BNNs. Figure \ref{dopingconcGlucose_AlGa} (left) shows the DOS of the G@Al-$h$-BNNs system for the three considered doping concentrations. As the doping concentration increases, the difference between the glucose LUMO and the BN sheet conduction band edge increases, thus decreasing the band gap from $3.6$ eV to $2.8$ eV, as shown in the inset. The charge transfer and the adsorption energy are almost independent of the dopant concentration. The increase in dopant concentration slightly increases the O-Al distance, Al-N bond length and N-Al-N angle. We also find that the doping concentration does not have a significant effect on the elongation of the C-O bond of glucose closest to the Al. Figure \ref{dopingconcGlucose_AlGa} (right) shows a qualitatively similar behavior for the G@Ga-$h$-BNNs case. 

\begin{figure}[H]
\begin{center}
\includegraphics[width=0.4\textwidth]{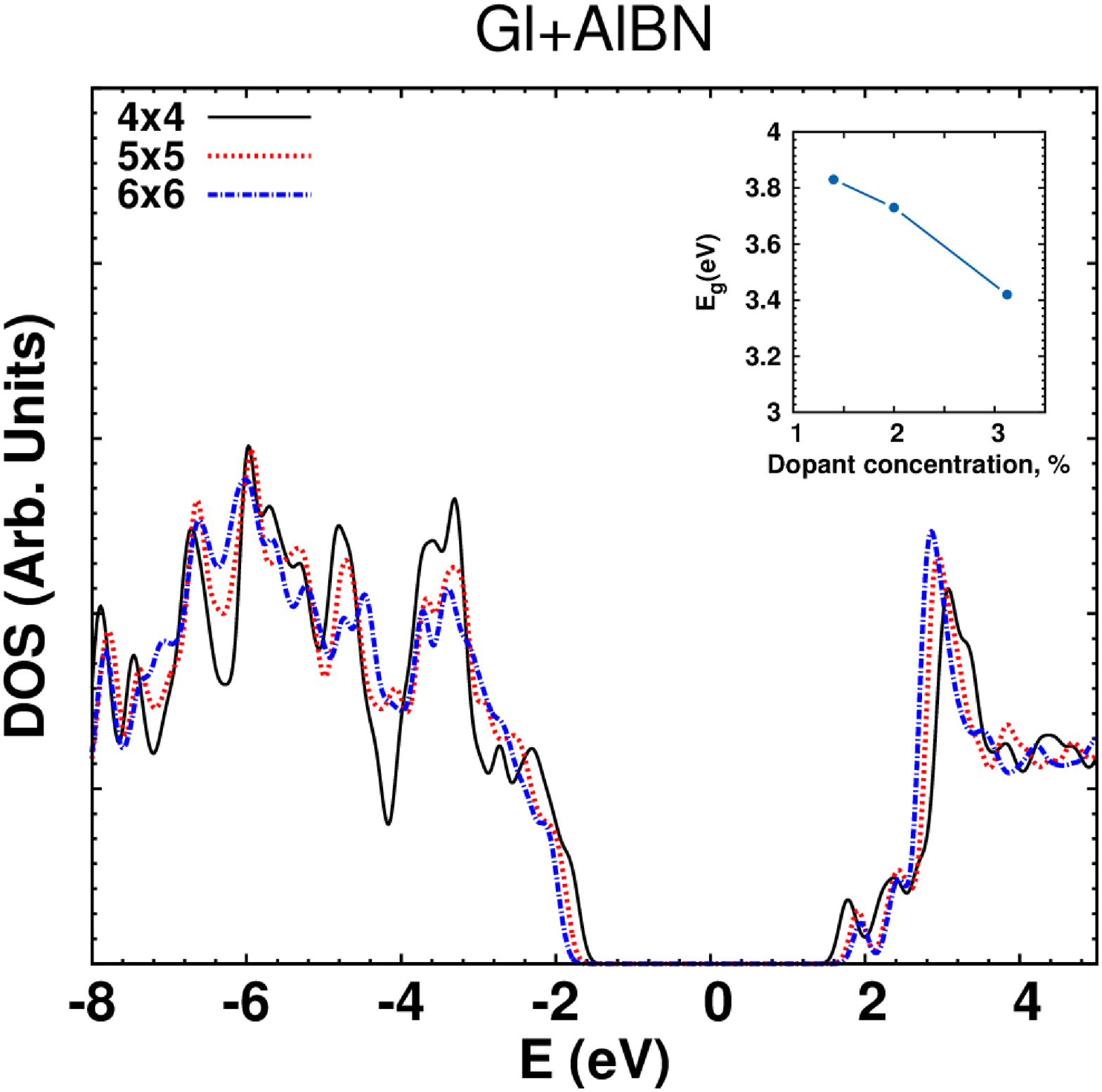}~\hspace{4mm}
\includegraphics[width=0.4\textwidth]{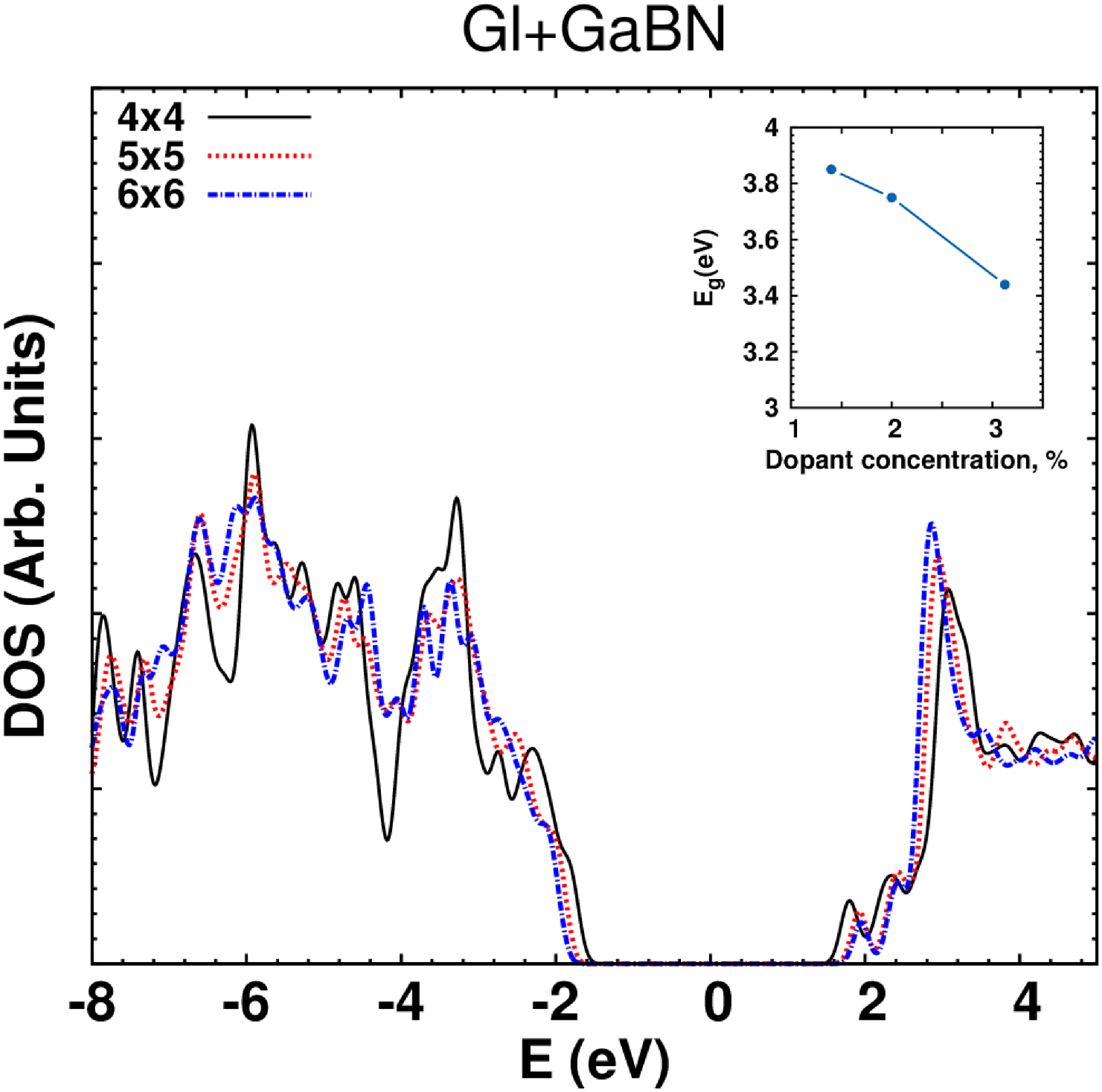}
\caption{(Color online) (Left) DOS of glucosamine adsorbed onto Al-doped $h$-BNNs, Gl@Al-$h$-BNNs, at various doping concentrations. (Right) DOS of glucosamine adsorbed onto Ga-doped $h$-BNNs, Gl@Ga-$h$-BNNs at various doping concentrations. Insets show the dependence of the energy gaps of both systems on the doping concentration.}
\label{dopingconcGlucosamine_AlGa}
\end{center}
\end{figure}

We next discuss the effect of doping concentration on the adsorption of glucosamine onto Al-$h$-BNNs and Ga-$h$-BNNs. Figure \ref{dopingconcGlucosamine_AlGa} (left) shows the DOS of the Gl@Al-$h$-BNNs system for the three doping concentrations. We notice a shift of the glucosamine LUMO to lower energy, which causes a decrease in the gap. The gap decreases with increasing doping concentration (as shown in the inset of Fig. \ref{dopingconcGlucosamine_AlGa} (left)), whereas the charge transfer between the glucosamine and the sheet, the adsorption energy, and structural details at the adsorption site are almost independent of the dopant concentration. The Gl@Ga-$h$-BNNs system follows a qualitatively similar behavior (Figure \ref{dopingconcGlucose_AlGa} (right)).

\subsection{van der Waals corrections} 

The approximate exchange-correlation functional can not give accurate adsorption energies \cite{Burke}. 
In the local density approximation (LDA) and the generalized gradient approximation (GGA), the long-range attractive dispersive interactions are not included. Many schemes have been employed to include dispersive corrections in DFT formulations \cite{Andersson,Dion,Tkatchenko}. The van der Waals density functional (vdW-DF) method describes the dispersive correction based
on the electron density, and has been used for studying various molecule-surface structures \cite{Sony,Chakarova}. Using 
the van der Waals DFT-D2 corrections \cite{grimme}, we find that the adsorption energy of glucose and glucosamine onto Al- and Ga- doped $h$-BNNs is increased by $\sim$ 0.6 eV, yielding a typical adsorption energy of about 2 eV. The van der Waals corrected results for various systems are shown in table \ref{resultstable}.

\begin{table}[H]
\caption{Adsorption energy using GGA ($E_{d}$) and van der Waals corrections ( $E_{d,\mbox{vdW}}$) for different concentrations}
\centering
{
\begin{tabular}{|c|c|c|c|c|c|c|c|}
\hline
Dopant & E&\multicolumn{3}{c|}{G onto doped sheet}&\multicolumn{3}{c|}{Gl onto doped sheet}
\\\hline
   &  &  $1.4$\% &  $2$\%&  $3$\% &  $1.4$\% &  $2$\%&  $3$\%  \\\hline
Al  &$E_{d}$&  1.21 & 1.27 & 1.28 & 1.41 &1.40&1.29 \\

 & $E_{d,\mbox{vdW}}$& 1.85 &  1.85 & 1.86 & 2.10 & 2.08&1.93\\\hline

Ga&$E_{d}$ &  0.70 & 0.78 & 0.79 & 1.05 & 1.04 &  0.94 \\
\raisebox{2ex}
& $E_{d,\mbox{vdW}}$&1.32 & 1.33& 1.32 & 1.71 & 1.70&1.58 \\\hline

\end{tabular}
}
\label{resultstable}
\end{table}

\section{Conclusion}

In conclusion, we have shown that the weak adsorption of glucose and glucosamine onto pristine boron nitride nanosheets can be significantly enhanced through the substitutional doping of the BN sheets with Al and Ga at the boron site. Glucose and glucosamine can adsorb onto these stable doped BN structures with an energy of the order of 1 eV. Adsorption to Al is found to be stronger than to Ga. The adsorption to Al- and Ga-doped $h$-BNNs leads to a change in their band gap. We find that increasing the doping concentration of Al and Ga decreases the band gap, but does not significantly affect the adsorption energy. By including the van der Waals corrections nearly doubles the adsorption energy. The results of our study may be utilized in the fabrication of biosensors, for applications where controlled adsorption of biomolecules is necessary, such as lab-on-a-chip, and for some catalysis applications.

\vspace*{-0.3cm}
\subsection*{Acknowledgment} 
\vspace*{-0.3cm}
This work has been done through a grant provided by Zewail City for Science and Technology. We acknowledge financial support from Center for Nanotechnology (CNT) and the Center for Fundamental Physics (CFP). The computations were performed on the Bibliotheca Alexandria supercomputer. 

\bibliography{g_al_bn}

\end{document}